\documentclass{article}

\usepackage{microtype}
\usepackage{graphicx}
\usepackage{subfigure}
\usepackage{booktabs} 

\usepackage{hyperref}
\usepackage{xurl}
\usepackage{tabularx}

\usepackage[accepted]{icml2025}

\usepackage{amsmath}
\usepackage{amssymb}
\usepackage{mathtools}
\usepackage{amsthm}

\usepackage[capitalize,noabbrev]{cleveref}

\theoremstyle{plain}

\theoremstyle{definition}

\theoremstyle{remark}

\usepackage[textsize=tiny]{todonotes}

\icmltitlerunning{AKD : Adversarial Knowledge Distillation For Large Language Models Alignment on Coding tasks}

\usepackage{graphicx}
\usepackage{booktabs}
\usepackage{xcolor}
\usepackage{float}
\usepackage{array}
\usepackage{parskip}
\usepackage{tcolorbox}
\tcbuselibrary{breakable}

\usepackage{listings}
\tcbuselibrary{listings, skins}
\newtcblisting{promptbox}{
    listing only,
    breakable,
    colback=black!5,       
    colframe=black,        
    listing options={
        basicstyle=\ttfamily\small,   
        keywordstyle=\color{blue},   
        breaklines=true,             
        tabsize=4                    
    },
    title=Prompt,                   
    fonttitle=\bfseries,            
    coltitle=black,                 
    boxrule=0.5mm                   
}

\begin{document}

\twocolumn[
\icmltitle{AKD : Adversarial Knowledge Distillation For Large Language Models Alignment on Coding tasks}



\icmlsetsymbol{equal}{*}

\begin{icmlauthorlist}
\icmlauthor{Ilyas Oulkadda}{yyy}
\icmlauthor{Julien Perez}{yyy}
\end{icmlauthorlist}

\icmlaffiliation{yyy}{Laboratoire de Recherche de l'EPITA (LRE), 14-16 rue Voltaire, 94270 Le Kremlin-Bicetre, France}

\icmlcorrespondingauthor{Julien Perez}{julien.perez@epita.fr}

\icmlkeywords{Machine Learning, ICML}

\vskip 0.3in
]

\begin{abstract}
The widespread adoption of Large Language Models (LLMs) for code generation, exemplified by GitHub Copilot\footnote{A coding extension powered by a Code-LLM to assist in code completion tasks} surpassing a million users, highlights the transformative potential of these tools in improving developer productivity. 
However, this rapid growth also underscores critical concerns regarding the quality, safety, and reliability of the code they generate. 
As Code-LLMs evolve, they face significant challenges, including the diminishing returns of model scaling and the scarcity of new, high-quality training data.
To address these issues, this paper introduces Adversarial Knowledge Distillation (AKD), a novel approach that leverages adversarially generated synthetic datasets to distill the capabilities of larger models into smaller, more efficient ones. 
By systematically stress-testing and refining the reasoning capabilities of Code-LLMs, AKD provides a framework for enhancing model robustness, reliability, and security while improving their parameter-efficiency. 
We believe this work represents a critical step toward ensuring dependable automated code generation within the constraints of existing data and the cost-efficiency of model execution.
\end{abstract}

\begin{figure*}[ht!]
    \centering
    \includegraphics[width=0.9\textwidth,keepaspectratio]{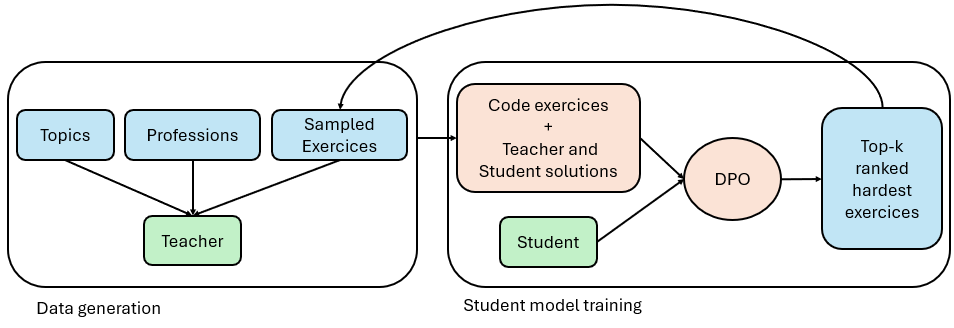}
    \caption{Adversarial Knowledge Distillation framework comprises three components, differentiated by color: models (green), training processes (red), and synthetic data generation elements (blue). In the initial step, topics and profession seeds are used to generate an initial dataset. Profession is acting as a persona to help diversify the dataset. For subsequent iterations, exercises are sampled based on margin rewards calculated from the previous training cycle. Once the exercises are selected or generated, the teacher model produces the "chosen" solutions, while the student model generates the "rejected" solutions. The resulting dataset, containing both chosen and rejected solutions, is then used to train the student model via Direct Preference Optimization (DPO). Margin rewards, which reflect the difference in performance between the teacher and student on specific tasks, are calculated to rank exercises. This ranking informs the sampling process for generating the next dataset, ensuring a targeted and iterative improvement cycle for the student model.}
\end{figure*}

\section{Introduction}

The rapid development and deployment of Large Language Models (LLMs), particularly those designed for code generation, such as GitHub Copilot, have started a transformative era in software development. 
These models hold the promise of significantly enhancing productivity by automating repetitive coding tasks and assisting developers with complex problem-solving. 
However, this evolution has also introduced pressing concerns regarding the quality, security, and ethical implications of the generated code. 
As developers increasingly depend on these models for critical applications, ensuring their outputs meet high standards of reliability and security has become imperative.

Traditional methods for aligning LLMs rely heavily on human-annotated datasets, which provide direct feedback and corrections. 
While effective, this approach faces challenges in scalability and adaptability due to the growing size of modern LLMs and the rapidly evolving nature of coding practices. 
The need for more sustainable and efficient alignment methodologies has become apparent, prompting research into alternative strategies.

On the other hand, generating datasets synthetically using LLMs can be beneficial in terms of scalability and efficiency \citep{Gunasekar2023TextbooksAA, Long2024OnLS}. 
However, it doesn't guarantee that the synthetic data will address all relevant cases. 
Synthetic data, generated in an open-ended manner, might lack the diversity and complexity of real-world scenarios, leading to gaps in the model's understanding and performance. 
This can result in models that perform well on synthetic benchmarks but struggle with real-world applications. 
Therefore, it's crucial to combine synthetic data generation with rigorous validation against real-world data to ensure the model's robustness and applicability.

To address these challenges, we propose Adversarial Knowledge Distillation (AKD), a novel framework designed to align Code-LLMs with functional and performance standards without the extensive reliance on human annotation. 
AKD integrates Direct Preference Optimization (DPO) \citep{rafailov2023direct} into a curriculum-driven adversarial dataset generation process \citep{gunasekar2023textbooks, openai2021asymmetric}. 
This approach enables distillation of knowledge \citep{hinton2015distillingknowledgeneuralnetwork} from a larger, more capable “teacher” model into a smaller, computationally efficient “student” model through an automatically defined curriculum by leveraging adversarially generated synthetic datasets of code exercices.

Concretely, AKD begins with the teacher model generating coding tasks, complete with function descriptions and solution outlines, which serve as a curated curriculum for the student model. 
Adversarial interactions are then used to stress-test the student model's reasoning and alignment capabilities, systematically exposing weaknesses and guiding improvements. 
The optimization process relies on DPO loss applied to the adversarial curriculum, allowing the student model to iteratively refine its alignment with the teacher’s distribution. 
Unlike conventional methods, such as Self-Supervised Fine-Tuning (SFT) based on next-token prediction, or DPO applied to unordered datasets, our approach leverages curriculum learning \citep{sukhbaatar2018intrinsic} to organize training data into a sequence that maximizes learning efficacy.

Preliminary results demonstrate the effectiveness of AKD in improving model performance and robustness. 
On the HumanEval \citep{chen2021evaluating} benchmarks, models trained with our adversarial curriculum exhibit higher reliability and security in generated code compared to those trained using traditional alignment methods. 
Furthermore, training logs reveal significant gains in reward margins, underscoring the effectiveness of adversarial datasets in enhancing the student model’s reasoning and adaptability.

This work seeks to advance the state of the art in aligning LLMs for automated code generation by addressing the critical challenges of quality, scalability, and ethical alignment and we aim to offer a scalable and sustainable pathway for developing smaller, more efficient models while ensuring their outputs remain secure, reliable, and aligned with industry standards.
In the following sections, we provide a review of related work in Section \ref{sec:soa}, introduce our adversarial adaptations of DPO in Section \ref{sec:method}, and analyze the performance of our approach in Section \ref{sec:xps}. 

\section{Related Work}
\label{sec:soa}

\textbf{Data Annotation}: Data annotation is still foundational for training and refining aligned Large Language Models (LLMs). 
Traditional methods often rely on manual annotation, which, while effective, faces significant scalability challenges as models and datasets grow in size and complexity. 
Manual annotation involves human annotators labeling data points, which can be time-consuming and costly, especially for large datasets.
As an example, \citep{li2023starcoder} employed 1,399 crowdworkers from 35 countries to annotate a dataset for identifying Personally Identifiable Information (PII) in source code. 
This highlights the extensive human effort required and the inherent difficulties in maintaining consistency and scalability across such diverse and large-scale annotation efforts. 
The process involved coordinating a large number of annotators, ensuring they understood the annotation guidelines, and managing the quality of the annotations. 
The challenges included variations in annotation quality due to differences in annotator expertise and cultural backgrounds, as well as the need for continuous monitoring and feedback to maintain consistency.
As LLMs require vast amounts of annotated data to achieve high performance, the traditional manual annotation approach becomes challenging. 
The need for efficient and scalable data production methods has led to the exploration of alternative approaches, such as semi-supervised learning, active learning, and automated annotation tools \citep{ruder2018semi, xie2020unsupervised, chen2020mixtext}. 
These methods aim to reduce the reliance on manual annotation by leveraging machine learning techniques to assist in the annotation process, thereby improving scalability and efficiency.

\textbf{Self-Instruct Approaches}: As a consequence to the expensive cost of annotation, leveraging larger models to autonomously generate high-quality, well-structured data has emerged as a scalable alternative to manual annotation. 
Notable examples include the phi-1.5 model ~\citep{Gunasekar2023TextbooksAA}, a 1.3B parameters model trained on high-quality, generated textbooks, and the \textit{Cosmopedia} dataset ~\citep{eldan2023tinystories}, which further demonstrates the utility of advanced models for data generation. 
The phi-1.5 model has been trained on synthetic "textbook-like" data, achieving performance comparable to models five times its size. 
The \textit{Cosmopedia} dataset, generated by the Mixtral-8x7B-Instruct-v0.1 model ~\citep{Jiang2024MixtralOE}, contains over 30 million files and 25B tokens, making it the largest open synthetic dataset to date. 
Moreover, incorporating topic sampling techniques enhances dataset diversity and ensures comprehensive coverage of scenarios, which is particularly critical for generating robust and adaptable models.

\textbf{Adversarial and Curriculum Learning}: Regarding automatic curriculum and dataset generation, adversarial learning has shown promise in various domains such as board games, video games, and robotics. 
This approach leverages the competitive nature of adversarial setups to generate increasingly complex task curricula, fostering progressive learning.
One notable technique is "asymmetric self-play" \citep{openai2021asymmetric}, where two agents, called Alice and Bob, engage in a competitive game. 
Alice is tasked with proposing challenging goals, while Bob aims to solve them. 
This dynamic leads to the discovery of highly diverse and complex goals without any human priors, thereby enhancing the learning process. 
For instance, in robotic manipulation tasks, asymmetric self-play has enabled the development of goal-conditioned policies that can generalize to previously unseen goals and objects \citep{openai2021asymmetric}. 
This method has been particularly effective in zero-shot generalization, where the learned policy can adapt to new tasks without additional training.
Another key technique is "intrinsic motivation and automatic curricula" \citep{sukhbaatar2018intrinsic}. 
This approach pits two versions of the same agent against each other, where one agent proposes tasks and the other attempts to complete them. 
Through an appropriate reward structure, the agents automatically generate a curriculum of exploration, enabling unsupervised training. 
This method has been successfully applied in environments that can be reset or are nearly reversible, allowing the agent to learn about its environment in an unsupervised manner. 
The generator network proposes tasks specified as goal states, and the agent learns to perform a wide set of tasks efficiently.
These insights motivates our adversarial training approach, where dynamically generated data progressively challenges the model. 
By mirroring a curriculum-learning framework tailored to Code-LLMs, we aim to enhance the model's ability to handle complex coding tasks. 
The adversarial setup ensures that the model is continually presented with new and challenging scenarios, fostering robust and adaptable learning. 
This approach can improves the model's performance on known tasks but also equips it with the ability to generalize to new, unseen tasks, making it more versatile and effective in real-world applications.

\textbf{Direct Preference Optimization and Knowledge Distillation}: Direct Preference Optimization (DPO) \citep{rafailov2023direct} represents a significant advancement in the alignment process of Large Language Models (LLMs). 
Unlike Proximal Policy Optimization (PPO), which rely on a reward model and can suffer from instability, DPO transforms the optimization challenge into a classification loss problem. 
This approach directly compares preferred and non-preferred outputs, streamlining the training process and reducing computational complexity. 
By doing so, DPO enhances also stability and makes the alignment of LLM outputs with human or teacher-generated preferences more efficient and effective. 
As the size of modern models continues to grow, concerns about their deployability for local use, especially for LLMs, have become more pronounced. Knowledge distillation \citep{hinton2015distillingknowledgeneuralnetwork} offers a compelling solution to this challenge. 
This technique enables smaller models to learn from the output (soft labels) of larger teacher models, effectively transferring knowledge while reducing the overall model size. 
By combining soft labels with hard-labeled annotated data during training, knowledge distillation ensures that the smaller model retains the performance benefits of the larger model without the computational overhead.
Our approach builds on the concept of knowledge distillation but adapts it specifically for adversarial fine-tuning scenarios. 
Instead of relying on hard labels, we focus on dynamic, adversarially generated datasets to guide the alignment of the student model. 
This adaptation not only simplifies the fine-tuning process but also ensures that the student model is exposed to a diverse and challenging set of examples, enhancing its robustness and adaptability. 
Knowledge distillation has been successfully applied in various domains, including natural language processing and computer vision, where it has helped create more efficient and deployable models. 
By leveraging these techniques, we aim to develop LLMs that are both high-performing and practical for real-world applications.


\section{Methodology}
\label{sec:method}

Adversarial Knowledge Distillation (AKD) introduces adversarial dynamics into the Direct Preference Optimization (DPO) algorithm. 
This is achieved by orchestrating an adversarial game between two models: a Teacher and a Student. 
The Teacher, being a larger and more proficient language model in the domain of coding, is tasked with generating coding exercises (prompts) and providing solutions that are assumed to align with expert human standards. 
These solutions are labeled as "chosen" answers. 
Conversely, the Student, a smaller model, attempts to solve the same exercises, and its solutions are labeled as "rejected" due to their presumed lesser alignment with expert preferences.

Our DPO setup involves a dataset with three key components: the prompt \( p \) (coding exercise), the chosen solution \( c \) (provided by the Teacher), and the rejected solution \( r \) (generated by the Student). 
This structured format facilitates the execution of DPO training iterations. 
To introduce adversariality, we conduct multiple rounds of these training iterations conditioned by the former errors of the Student model. 
At the end of each iteration, we calculate the loss for each training sample, enabling us to determine the rewards for each instance.

The loss function for a given sample can be defined as:
\[ L(x, c, r) = -\log \left( \frac{\exp(R(c, x))}{\exp(R(c, x)) + \exp(R(r, x))} \right) \]
where \( R(c) \) and \( R(r) \) are the rewards for the chosen and rejected solutions, respectively.
\subsection*{Adversarial Step}

The core of the adversariality in our framework lies in how we identify and target the weaknesses of the Student model. A critical metric in this setup is the \textit{margin reward}, which measures the difference between the rewards assigned to the chosen and rejected answers. The margin reward \( M \) is computed as:
\[
M = R(c) - R(r)
\]
where \( R(c) \) and \( R(r) \) represent the rewards for the chosen and rejected answers, respectively.

A small margin indicates that the Student model struggles to differentiate between the correct and incorrect solutions. This forms the foundation of the adversariality: we aim to focus on these challenging areas to push the model's boundaries.

To operationalize this, we apply a softmax over the negative margins, which transforms the margin rewards into a probability distribution. Prompts with lower margins (indicating greater difficulty) are assigned higher probabilities, ensuring that these weaknesses are emphasized in subsequent training. The probability \( P(p) \) of selecting a prompt \( p \) is defined as:
\[
P(p) = \frac{\exp(-M(p))}{\sum_{p' \in B} \exp(-M(p'))}
\]
where \( B \) represents the batch of prompts.

This adversarial sampling process occurs at the batch level, rather than at the level of individual samples. This approach not only diversifies the topics and challenges presented to the Student model but also ensures that the dataset adapts dynamically to target its weaknesses. The iterative nature of this process drives the adversariality by continually generating and focusing on data that challenges the model, fostering robust learning.

The final step involves using these selectively sampled exercises to generate new prompts targeting the identified weaknesses of the Student model. 
Initial attempts to generate similar exercises directly from the hardest samples resulted in adversarial datasets that were smaller and overly similar to the initial dataset. 
To address this issue, we implemented three distinct prompting strategies to diversify and enrich the generated adversarial exercises:

\textbf{Incremental Approach:} This strategy gradually increases the difficulty of the exercises, ensuring that the Student model is continually challenged but not overwhelmed. By starting with the hardest samples identified in the previous iteration and incrementally increasing the difficulty, we aim to create a smooth learning curve. The difficulty level \( D \) of a new exercise is given by:
\[ D_{\text{new}} = D_{\text{hard}} + \Delta D \]
where \( D_{\text{hard}} \) is the difficulty of the hardest sample and \( \Delta D \) is a small increment. 
This incremental approach helps the Student model to build on its existing knowledge and skills, progressively enhancing its capabilities. 
The small increments ensure that the model is exposed to new challenges that are just beyond its current abilities, promoting effective learning and adaptation.

\textbf{Opposite Approach:} This approach is motivated by the need to challenge the Student model's assumptions and biases. 
By generating exercises that appear similar to familiar tasks but introduce unexpected variations, we force the model to adapt to new and unpredictable scenarios. 
The similarity \( S \) between the new exercise and the original exercise is maintained within a certain threshold:
\[ S(p_{\text{new}}, p_{\text{original}}) \leq \theta \]
where \( \theta \) is the similarity threshold. 
This strategy ensures that the model does not rely on superficial patterns or shortcuts but instead develops a deeper understanding of the underlying principles. 
By confronting the model with exercises that defy its expectations, this approach encourages it to generalize better and become more robust to diverse and unseen situations.

\textbf{Deceptive Approach:} This strategy focuses on creating exercises that appear deceptively simple but conceal underlying complexity. 
The goal is to challenge the Student model's ability to discern and handle hidden intricacies. 
The complexity \( C \) of the exercise is hidden behind a deceptive simplicity:
\[ C(p_{\text{deceptive}}) > C(p_{\text{simple}}) \]
where \( p_{\text{simple}} \) is a straightforward exercise. 
This approach is particularly effective in training the model to avoid overconfidence and to develop a more nuanced understanding of the tasks. 
By presenting exercises that require deeper analysis and careful consideration, we promote the model's ability to handle subtle and complex problems, enhancing its overall problem-solving skills.

These diversified strategies ensure that the adversarial datasets are not only richer in variety but also more robust in challenging the Student model across different dimensions. 
The resulting iterative process dynamically adapts to the evolving capabilities of the model. 
As the Student model improves, the Teacher model continually generates new challenges that push the boundaries of the Student's current abilities. 
This adaptive mechanism ensures that the training remains consistently robust and diverse, preventing the model from plateauing or overfitting to specific types of exercises. 
By continually introducing new and varied challenges, we foster a learning environment that promotes sustained growth and improvement.
All training details, including the exact prompts used for dataset generation, the parameters for each prompting strategy, and the iterative training schedule, are provided in the Appendix.

\section{Experiments}
\label{sec:xps}

In this section, we address research questions related to the performance and effectiveness of our proposed method. 
Traditionally, knowledge distillation combines soft labels, i.e. probability distributions produced by the teacher model, and hard labels, i.e. ground truth labels, to train a student model from scratch. 
However, due to computational constraints, our experiments focus on fine-tuning pre-trained models rather than training from scratch. 
In our approach, the teacher's soft labels are treated as hard labels for fine-tuning. 
Despite this limitation, we hypothesize that our method remains effective as a fine-tuning strategy, offering performance gains through adversarially ordered training steps. 

We evaluate our approach using two widely-adopted code generation benchmarks: HumanEval~\citep{chen2021evaluating} and MBPP~\citep{austin2021programsynthesislargelanguage}. To ensure rigorous assessment of functional correctness, we employ the EvalPlus framework~\citep{evalplus}, evaluating on both the original benchmarks and their enhanced EvalPlus+ variants. These augmented versions retain the original problem statements while substantially expanding the test suites—HumanEval+ adds an average of 80× more tests per problem compared to HumanEval, and MBPP+ provides 35× more tests than MBPP~\citep{evalplus}. This comprehensive evaluation strategy enables us to measure both base performance and robustness against subtle implementation errors.

\begin{table}[t]
\caption{Models used in our distillation experiments.}
\label{tab:models}
\centering
\small
\begin{tabularx}{\columnwidth}{@{}lXX@{}}
\toprule
\textbf{Model} & \textbf{Context Length} & \textbf{Size} \\
\midrule
Llama-3.1-8B  & 128K & 8B \\
Llama-3.2-1B  & 128K & 1B \\
Phi-1.5  & 4500 & 1.3B \\
DeepSeek-R1-Qwen-7B  & 4500 & 7B \\
DeepSeek-R1-Llama-8B  & 130K & 8B\\
Qwen2.5-7B  & 4500 & 7B \\
\bottomrule
\end{tabularx}
\vspace*{-2mm} 
\end{table}

Our distillation framework is evaluated across contemporary open-source models spanning 1B to 8B parameters, including Meta's general-purpose Llama-3.1-8B and Llama-3.2-1B \citep{grattafiori2024llama3herdmodels}, DeepSeek's reasoning-optimized R1 series (DS-R1-Qwen-7B and DS-R1-Llama-8B) \citep{deepseekai2025deepseekr1incentivizingreasoningcapability}, Qwen's code-specialized Qwen2.5-Coder-7B \citep{hui2024qwen25codertechnicalreport}, and Microsoft's compact Phi-1.5. This selection encompasses both general instruction-following architectures and domain-specific variants, with more details provided in Table~\ref{tab:models}. The spectrum of model sizes and architectural specializations enables comprehensive analysis of knowledge transfer efficacy across different capability profiles.

The rest of section is organized into three subsections. 
First, we evaluate whether our method outperforms self-supervised fine-tuning approaches. 
Second, we assess the contribution of adversarial training steps to model performance. 
Finally, we explore the applicability of our teacher-student framework for speculative decoding \citep{leviathan2023fastinferencetransformersspeculative}, focusing on its impact on speed and accuracy in text generation. 
These experiments aim to validate the capabilities of AKD across multiple dimensions, demonstrating its effectiveness in improving fine-tuning, enhancing alignment, and supporting efficient decoding strategies.

Since we were limited in resources, we were not able to run AKD at larger scale generating larger datasaets. Our experiments are done on rather small datasets varying from 700k tokens to 3M tokens using Llama 3 tokenizer. We believe that the results could be scaled up with dataset size as seen with the DS-R1 and Llama 3.2 1B pair showing higher performance with a dataset size of 3M tokens. However, we also notice high improvements on the Qwen2.5-Coder 7B and Phi 1.5 pair with only 700k tokens.

\begin{figure}
    \centering
    \includegraphics[width=1\linewidth]{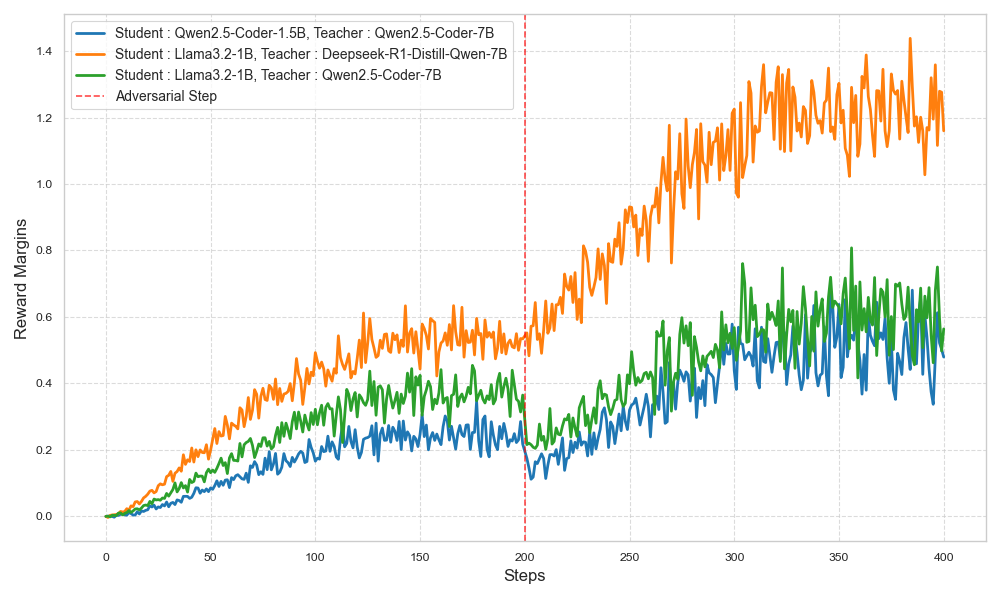}
    \caption{Margin Rewards during training: At the 200-step mark, an adversarial step generates a new dataset, continuing training with the same optimizer states. A small performance dip is observed in some experiments due to increased exercise difficulty.}

    \label{fig:enter-label}
\end{figure}

\begin{table*}[ht]
\centering
\renewcommand{\arraystretch}{1.3}
\setlength{\tabcolsep}{5pt}  
\begin{tabular}{@{}llc@{\hspace{8pt}}cc@{\hspace{8pt}}cc@{\hspace{8pt}}cc@{\hspace{8pt}}cc@{}}
\toprule
\textbf{Teacher} & \textbf{Student} & \textbf{Size} & 
\multicolumn{2}{c}{\textbf{HumanEval}} & 
\multicolumn{2}{c}{\textbf{HumanEval+}} & 
\multicolumn{2}{c}{\textbf{MBPP}} & 
\multicolumn{2}{c}{\textbf{MBPP+}} \\
\cmidrule(lr){4-5} \cmidrule(lr){6-7} \cmidrule(lr){8-9} \cmidrule(lr){10-11}
 & & (M) & T & S ($\Delta$) & T & S ($\Delta$) & T & S ($\Delta$) & T & S ($\Delta$) \\
\midrule
DeepSeek-R1-Qwen-7B & Llama-3.2-1B & 3.0 & 39.6 & 36.5\,(\textbf{+3.6}) & 37.8 & 29.2\,(0.0) & 43.1 & 41.5\,(\textbf{+8.7}) & 37.3 & 34.9\,(\textbf{+6.9}) \\

Llama-3.1-8B & Phi-1.5 & 6.4 & 69.5 & 39.0\,(+3.1) & 62.8 & 31.0\,(+1.0) & 68.3 & 53.9\,(-1.1) & 54.8 & 43.9\,(-0.8) \\

Qwen2.5-Coder-7B & Phi-1.5 & 1.1 & 61.6 & 38.2\,(+2.3) & 53.0 & 31.7\,(+1.7) & 76.9 & 52.9\,(-2.1) & 62.9 & 43.3\,(-1.1) \\

Qwen2.5-Coder-7B & Llama-3.2-1B & 0.7 & 61.6 & 34.7\,(+1.8) & 53.0 & 32.3\,(\textbf{+3.2}) & 76.9 & 32.5\,(-0.3) & 62.9 & 28.0\,(0.0) \\

Llama3.1-8B & Llama-3.2-1B & 1.7 & 69.5 & 30.4\,(-2.5) & 62.8 & 28.0\,(-1.2) & 68.3 & 34.7\,(+1.9) & 54.8 & 29.9\,(+1.9) \\

\bottomrule
\end{tabular}
\caption{Knowledge Distillation Performance. Size column indicates millions of tokens in the training dataset. T: Teacher score, S: Student score with improvement delta ($\Delta$) from initial version. The data we generate is mainly in the HumanEval format however a small portion is reserved for natural language instruction similar to MBPP format. Constant improvements are seen on HumanEval except for the Llama pair, where the 1B model is already distilled from larger counterparts.}
\label{tab:distillation_results}
\end{table*}

\subsection{Experimental settings}

For data generation, detailed in Table ~\ref{tab:hp}, the teacher model uses a longer maximum sequence length only for exercise creation, while solutions use the same length as the student model to avoid bias and ensure fair output sizes. To optimize memory efficiency during training, we employ gradient accumulation of 4 and gradient checkpointing, with a low learning rate reflecting a later training stage post-pretraining, instruction fine-tuning, and alignment. The framework performs best initially but degrades with more repetitions, so we limit repetitions. Future work could explore this with larger datasets and more complex tasks. We use LoRA \citep{hu2021lora}  for fine-tuning, targeting attention projection matrices and MLP layers to enhance efficiency and performance, the hyperparameters are detailed in Table ~\ref{tab:lora}.

\begin{table}[h!]
\centering
\begin{tabular}{@{}ll@{}}
\toprule
\textbf{Parameter}                     & \textbf{Value}              \\ \midrule
Number of subtopics per topic          & 10                          \\
Number of exercises per subtopic       & 10                          \\
Oracle temperature                     & 1                           \\
Oracle max length                      & 2048                        \\
Student temperature                    & 0.6                         \\
Student max length                     & 512                         \\
Number of steps                        & 150                         \\
Per device train batch size            & 4                           \\
Learning rate                          & 5e-6                        \\
Beta                                   & 0.01                        \\
Repetitions                            & 5                           \\ 
\midrule
Quantization                                   & BF16                        \\
Gradient accumulation steps            & 4                           \\
Learning rate scheduler type           & cosine                      \\
Optimizer                              & Paged AdamW         \\
\bottomrule
\end{tabular}
\caption{Main Hyperparameters for Model Training and Data Generation.}
\label{tab:hp}
\end{table}

\begin{table}[h!]
\centering
\begin{tabular}{@{}ll@{}}
\toprule
\textbf{LoRA Configuration}            & \textbf{Value}              \\ \midrule
Rank (r)                               & 16                          \\
LoRA alpha                             & 32                          \\
LoRA dropout                           & 0.05                        \\
Bias                                   & none                        \\
Target modules                         & \{q\_proj, k\_proj, v\_proj, o\_proj, \\ 
                                       & gate\_proj, up\_proj, down\_proj\} \\ \bottomrule
\end{tabular}
\caption{Configuration Parameters for LoRA adapters.}
\label{tab:lora}
\end{table}

\subsection{Performance Comparison: Proposed Framework vs. Self-Supervised Fine-Tuning}

We compare the performance of AKD against traditional self-supervised fine-tuning approaches. Self-supervised methods rely on the model's ability to learn from the data itself without external supervision. While these methods have shown promise, they often struggle with capturing the nuanced preferences and complexities that a teacher model can provide. By leveraging the teacher's soft labels as hard labels, AKD aims to bridge this gap, offering a more guided and effective fine-tuning process. We evaluate the performance of both methods on a variety of benchmarks to determine if AKD provides a significant advantage.

Concretely, we evaluate the performance of Adversarial Knowledge Distillation (AKD) against a baseline self-supervised fine-tuning (SFT) method using next-token prediction loss. Both methods are benchmarked on the HumanEval dataset, with AKD demonstrating unique advantages.
The AKD framework creates a preference dataset by generating coding questions and solutions using a teacher model. These questions are then attempted by the student model, and the resulting answers are paired to form a preference dataset, where the teacher's response is marked as the preferred solution. The training process uses a contrastive loss function to align the student’s predictions closer to the teacher’s distribution, maximizing learning efficiency.

For the baseline, we explore two approaches to self-supervised fine-tuning. The first approach uses the same synthetic dataset generated by the teacher for AKD, where the questions and teacher solutions are concatenated to create a training corpus. The second approach fine-tunes the student model on the APPS dataset, a challenging coding benchmark comprising approximately 5,000 samples. While both approaches yield equivalent performance on HumanEval, significant insights emerge from the dataset comparison.

Table~\ref{tab:framework_vs_selfsupervised} shows that AKD achieves a comparable 38\% accuracy on the HumanEval benchmark, matching the performance of SFT on the APPS dataset. Notably, the AKD-generated dataset, consisting of only 1.6k samples, achieves the same results as the handcrafted APPS dataset, underscoring the quality and efficiency of AKD's dataset generation process.
These findings highlight that AKD not only matches the performance of traditional self-supervised fine-tuning on a significantly smaller dataset but also offers a scalable and automated approach to creating high-quality datasets for coding tasks. Future work will explore scaling AKD to additional domains and larger datasets while maintaining its competitive performance.

\begin{table*}[ht]
\centering
\begin{tabular}{>{\raggedright\arraybackslash}p{4cm} >{\centering\arraybackslash}p{3cm} >{\centering\arraybackslash}p{3cm}}
\toprule
\textbf{Method} & \textbf{Accuracy (HumanEval, \%)} & \textbf{Dataset Size (Samples)} \\
\midrule
Self-Supervised Fine-Tuning (AKD dataset) & 38 & 1.6k \\
Self-Supervised Fine-Tuning (APPS train) & 38 & 5k \\
Adversarial Knowledge Distillation & 38 & 1.6k \\
\bottomrule
\end{tabular}
\caption{Comparison of Benchmark Performance Between Adversarial Knowledge Distillation and Self-Supervised Fine-Tuning. Evaluated on Qwen2.5 Coder 7B / Llama 3.2 1B}
\label{tab:framework_vs_selfsupervised}
\end{table*}

\subsection{Benefits of Adversarial Training Steps}

Here, we evaluate the specific contributions of adversarial training steps to the overall performance of the model. Adversarial training introduces a dynamic element to the learning process, continually challenging the model with increasingly difficult tasks. This approach not only helps in identifying and addressing the model's weaknesses but also promotes robustness and adaptability. We analyze the performance metrics at different stages of the adversarial training process to quantify the benefits of this approach. By comparing models fine-tuned with and without adversarial steps, we aim to demonstrate the tangible improvements brought by our method.
Unlike standard training, where the entire dataset is presented in random order, our method introduces an iterative approach. Exercises are sequentially presented with increasing difficulty, leveraging a curriculum designed to maximize the student model's learning efficiency.

To assess the impact of this strategy, we compare our framework to a baseline DPO approach. In the DPO baseline, the entire synthetic dataset is concatenated and presented as a single batch during training. By contrast, our adversarial approach incrementally adapts the dataset, using performance feedback from the student model to rank and order the exercises. This ranking ensures that the student encounters progressively more challenging tasks, simulating a teacher-like guidance mechanism.

As shown in Table~\ref{tab:adversarial_vs_dpo}, our adversarial framework consistently outperforms the DPO baseline, achieving a 4 percentage point improvement in accuracy. This result highlights the importance of gradual difficulty progression, which appears to enhance the student model’s learning dynamics and overall performance. Furthermore, this finding aligns with prior research on curriculum learning, suggesting that structured training paradigms can yield significant gains in model efficacy.

\begin{table*}[ht]
\centering
\begin{tabular}{lcc}
\toprule
\textbf{Method} & \textbf{Accuracy (HumanEval, \%)} & \textbf{Improvement (\%)} \\
\midrule
DPO Baseline & 35 & -- \\
Adversarial Training & \textbf{38} & +3 \\
\bottomrule
\end{tabular}
\caption{Performance Comparison: Adversarial Training vs. DPO Baseline}
\label{tab:adversarial_vs_dpo}
\end{table*}

\subsection{Speculative Decoding with Teacher-Student Framework}

Finally, we explore the potential of our teacher-student framework in the context of speculative decoding. Speculative decoding is a technique aimed at accelerating text generation by predicting future tokens based on current predictions. This method relies heavily on the accuracy and speed of the model's predictions. We investigate how our AKD framework can enhance speculative decoding by providing more accurate and efficient predictions. By evaluating the speed and accuracy of text generation with and without our framework, we assess the practical benefits of integrating AKD with speculative decoding techniques.

In this subsection, we explore the applicability of our framework for speculative decoding, leveraging the teacher as the main model and the student as the assistant. Speculative decoding is a decoding strategy that accelerates text generation by delegating initial token predictions to a smaller assistant model and verifying these predictions with the main model. The primary objective is to benchmark the accuracy of this setup relative to the number of forward passes required on the main model, thereby evaluating the trade-off between generation speed and performance.

To achieve effective speculative decoding, the assistant model must satisfy two key criteria. Firstly, it should be smaller in size to ensure computational efficiency. Secondly, and most importantly, the output token distribution of the assistant must closely resemble that of the main model. A higher degree of similarity minimizes the number of forward passes to the main model, as fewer corrections are required for the assistant's predictions. 

While speculative decoding is traditionally implemented using distilled or quantized versions of the main model, these approaches present challenges: distillation of an entire model remains computationally expensive, and quantization introduces complexities in ensuring compatibility and maintaining performance. Our adversarial knowledge distillation (AKD) framework was proposed as an alternative to generate assistant models tailored for speculative decoding by fine-tuning the student model on datasets generated by the teacher.

However, our experiments did not demonstrate improvements in speculative decoding performance using teacher-student pairs trained within the same model family (e.g., Llama 8B as the teacher and Llama 1B as the student). Speculative decoding requires both the assistant and main models to have compatible tokenizers, which restricted our experiments to models from the same family. This limitation likely contributed to the lack of improvement, as models within the same family are typically trained on overlapping datasets. Consequently, the teacher's generated text closely resembles the pretraining data, leaving limited room for the student to learn beyond the shared knowledge already captured during pretraining. 

We theorize that this limitation is due to the inherent saturation of knowledge within a single model family. Since both teacher and student models are trained on similar corpora, the teacher’s prompts and solutions do not introduce novel challenges for the student. This lack of novelty restricts the student's ability to fine-tune its token distribution to better align with the teacher, which is critical for successful speculative decoding. 

While our results did not confirm the efficacy of AKD for speculative decoding in this constrained setup, we hypothesize that using models from different families with distinct training data but same tokenizers may yield better results. Future work should explore cross-family speculative decoding setups, where the teacher and student are chosen from different architectures, enabling richer adversarial interactions and more diverse datasets for fine-tuning the student.

\section{Conclusion}

In this paper, we have introduce Adversarial Knowledge Distillation, a novel alignment framework for language models, with a focus on code-focused LLMs. 
AKD demonstrates promise as a flexible, task-agnostic method adaptable to a variety of domains. Our experiments revealed that AKD can yield meaningful performance improvements with limited data, outperforming traditional methods such as standard fine-tuning and Direct Preference Optimization in scenarios where pre-existing datasets are unavailable.
Through the use of synthetic datasets comprising approximately 1400 samples, we observed that AKD enables student models to improve relative to their teachers in code-related tasks. 
This underscores its potential as an efficient approach to aligning models without the need for large curated datasets. Scaling the dataset size and diversifying its themes represent promising directions to further enhance AKD’s effectiveness.
In the future, methods like AKD could be scaled and adapted to other well-defined tasks, leveraging its task-agnostic nature. 
This flexibility makes it applicable beyond code, paving the way for broader utility across domains.
For code-specific applications, additional avenues of improvement could be explored by building on the current work. One promising direction is integrating compiler feedback to refine the quality of synthetic datasets. Compiler feedback could serve as a valuable signal to ensure datasets align closely with task requirements, enhancing model performance for coding challenges.

\bibliographystyle{icml2025}   

\newpage
\nocite{*}
\newpage

\section{Appendix}

This appendix provides detailed prompt templates for generating synthetic Python programming exercises. These prompts are designed to create diverse and challenging exercises for learners, ranging from fundamental to advanced levels. Each category serves a specific purpose in dataset generation, ensuring a mix of conceptual understanding and practical application.

\subsection{Topics Generation}
This prompt assists in generating structured subtopics for a given Python concept. The goal is to ensure a logical progression of topics, covering both fundamental and advanced aspects. The generated subtopics should be distinct, build upon each other, and allow for the creation of multiple types of exercises, including theory, practice, and debugging.

\begin{promptbox}
As a Python textbook author, create {n} distinct subtopics for the main topic '{topic}'.

Requirements:
- Each subtopic should be broad enough to generate multiple exercise types (theory, practice, debugging).
- Include both fundamental and advanced concepts.
- Ensure topics build on each other in a logical learning progression.
- Mix conceptual and practical application areas.
- Avoid overlapping subtopics.

Format your response as a Python list of strings, like this:
['Basic String Operations', 'String Formatting Methods', 'Regular Expressions']

Return only the Python list with no additional text or explanation.
\end{promptbox}

\subsection{Initial Synthetic Dataset Generation, Code Completion Version}
This prompt generates Python code completion exercises tailored to a specific topic. Each exercise is framed within a real-world professional context to enhance engagement and relevance. The structure ensures that learners focus on writing functional code while adhering to given constraints.

\begin{promptbox}
Create {n} unique and challenging Python code completion exercises on the topic of "{topic}".
Each exercise should be framed in the context of a {profession}'s work to make it more engaging.

Structure each exercise as follows:

```
def function_name(parameters):
    """
    Description of the task or problem to solve, providing all necessary context.
    """
    # Solution code starts here (in Python)
```

Guidelines:
- Avoid using classes
- Make exercises challenging
\end{promptbox}

\subsection{Initial Synthetic Dataset Generation, Natural Language Version}
This prompt generates Python programming exercises described in natural language. The exercises are embedded within realistic professional scenarios, ensuring domain relevance. Each exercise includes problem statements, input-output specifications, test cases, and constraints to guide learners.

\begin{promptbox}
Create {n} unique and challenging Python programming exercises on the topic of "{topic}",
framed within the context of a {profession}'s daily work scenarios.

Structure each exercise as follows:

[PROBLEM]
Problem: [Natural language description that presents the programming challenge in a {profession}-related scenario]

Input: [Clearly specify the input format using domain-specific examples]

Output: [Clearly specify the expected output format]

Examples:
[Provide 2-3 test cases with profession-relevant data and explanations]

Constraints:
[State any constraints or special considerations]
[PROBLEM]

Example scenario structure:
- For a chef: "Given a list of recipe preparation times, find the optimal cooking schedule..."
- For an architect: "Given dimensions of building materials, calculate the most efficient arrangement..."

Guidelines:
- Use profession-specific terminology and realistic scenarios
- Make the problem mathematically sound while maintaining professional context
- Include domain-relevant example data in test cases
- Do NOT provide the solution, only the problem description
- Make SURE to write the problem between the [PROBLEM] tokens
\end{promptbox}

\subsection{Adversarial Dataset Generation, Incremental Approach}
This prompt generates exercises that incrementally increase complexity based on a given reference problem. These exercises introduce additional edge cases and nuanced difficulties, encouraging deeper problem-solving skills.

\begin{promptbox}
Based on the reference exercise '{reference}', generate {n} new coding exercises that introduce additional edge cases and increased complexity.

Requirements:
- Build upon the original task, adding nuanced complexity.
- Avoid using classes, but require complex logic and multiple edge cases.
- Provide a solution following the exercise.

Format:
```
def function_name(parameters):
    \"\"\"Exercise description with additional complexity.\"\"\"
    # Solution code here
```
\end{promptbox}

\subsection{Adversarial Dataset Generation, Opposite Approach}
This prompt creates exercises that challenge the assumptions of a given reference problem. The aim is to encourage adaptability in solving unexpected variations of a familiar topic.

\begin{promptbox}
    
Create {n} adversarial coding exercises for the topic related to '{reference}' that challenge conventional assumptions made in the original exercise.

Requirements:
- Shift expected assumptions, requiring the model to adapt to unexpected scenarios.
- Keep exercises challenging and avoid using classes.
- Provide solutions immediately after each exercise.

Format:
```
def function_name(parameters):
    """Exercise with altered assumptions."""
    # Solution code here
```
\end{promptbox}

\newpage
\subsection{Adversarial Dataset Generation, Deceptive Complexity Approach}
This prompt generates exercises that appear simple at first glance but contain hidden complexities. These problems test the learner's ability to identify and handle underlying difficulties in problem-solving.

\begin{promptbox}
Generate {n} coding exercises inspired by '{reference}' that appear simple but involve hidden complexities.

Requirements:
- Exercises should look straightforward but require complex solutions.
- Avoid classes and focus on deceptive problem setups.
- Solutions should immediately follow the exercise.

Format:
```
def function_name(parameters):
    """Exercise with hidden complexities."" 
    # Solution code here
```    
\end{promptbox}

\end{document}